\documentclass[runningheads]{llncs}
\usepackage[T1]{fontenc}
\usepackage{graphicx}
\usepackage{amsmath}
\usepackage{hyperref}
\usepackage{multirow}
\usepackage{makecell}
\usepackage{comment}
\usepackage{subcaption}
\usepackage{caption}
\usepackage{floatrow}
\usepackage{booktabs}  
\usepackage{rotating}
\newcommand\numberthis{\addtocounter{equation}{1}\tag{\theequation}}

\newcommand{\expand}[1]{$\quad$#1$\quad$}

\begin{document}
%
\title{On the Reproducibility of Learned Sparse Retrieval Adaptations for Long Documents}
\titlerunning{Rep. of LSR Adaptations for Long Documents}
%

\author{Emmanouil Georgios Lionis\orcidID{0009-0004-3931-9657} \and
Jia-Huei Ju \orcidID{0000-0003-2247-3370}}

\authorrunning{Lionis and Ju}
%
\institute{University of Amsterdam, The Netherlands\\ 
\email{akis.lionis@student.uva.nl}, \email{j.ju@uva.nl}
}
\maketitle              
\begin{abstract}
Document retrieval is one of the most challenging tasks in Information Retrieval. It requires handling longer contexts, often resulting in higher query latency and increased computational overhead. 
Recently, Learned Sparse Retrieval (LSR) has emerged as a promising approach to address these challenges.
Some have proposed adapting the LSR approach to longer documents by aggregating segmented document using different post-hoc methods, including n-grams and proximity scores, adjusting representations, and learning to ensemble all signals. In this study, we aim to reproduce and examine the mechanisms of adapting LSR for long documents. 
Our reproducibility experiments confirmed the importance of specific segments, with the first segment consistently dominating document retrieval performance. 
Furthermore, We re-evaluate recently proposed methods -- ExactSDM and SoftSDM -- across varying document lengths, from short (up to 2 segments) to longer (3+ segments). 
We also designed multiple analyses to probe the reproduced methods and shed light on the impact of global information on adapting LSR to longer contexts.
The complete code and implementation for this project is available at: \href{https://github.com/lionisakis/Reproducibilitiy-lsr-long}{https://github.com/lionisakis/Reproducibilitiy-lsr-long}.
\end{abstract}

\section{Introduction}

Research on document retrieval has become a prominent focus in neural Information Retrieval (IR)~\cite{dai2019deeper,zhang2021comparing}.
However, the self-attention mechanism in transformers increases computational complexity, scaling quadratically with the input sequence length. As a result, these models are typically constrained to a fixed input length, such as 512 tokens in the case of BERT \cite{devlin2018bert}.
An intuitive approach to addressing this limitation is to divide long documents into multiple segments\footnote{A segment refers to a text unit limited to 512 tokens, representing a shorter portion of a full document. The terms ``segment'' and ``passage'' are used interchangeably throughout.}, encode each segment individually, and then aggregate the resulting scores or representations to evaluate the document's overall relevance \cite{boytsov2022understanding,li2023parade,macavaney2019cedr,zhang2021comparing}. 
While in traditional lexical approaches, such as BM25, dependencies between terms can be incorporated with Sequential Dependence Model (SDM) model~\cite{metzler2005markov}, which extends the traditional bag-of-words framework~\cite{bendersky2010learning,huston2014comparison,metzler2005markov}. 
Recently, SDM model has been applied to learned sparse retrieval approaches (LSR)~\cite{formal2021splade} by Nguyen et al.~\cite{nguyen2023adapting}. 
They integrate lexical features into sparse bi-encoder models and introduce ExactSDM and SoftSDM for document retrieval tasks.

To better understand the potential of LSR on document retrieval, we reproduce the study ``Adapting Learned Sparse Retrieval for Long Documents'' by Nguyen et al. \cite{nguyen2023adapting}. 
We aim to compare the effectiveness of diverse segment aggregation strategies (e.g., representation or score) and particularly validate the proposed adaptation of ExactSDM and SoftSDM with sparse encoders.
We are going to verify the following claims from the original paper:
\begin{itemize}
    \item[] \textbf{Claim 1:} \label{Claim_1} Considering more segments in a document, the performance of representation aggregation decreases, whereas Score-Max remains unaffected.
    \item[] \textbf{Claim 2:} \label{Claim_2} The ExactSDM and SoftSDM models improve LSR performance through proximity-based matching, with ExactSDM emerging as the most effective aggregation method for targeted benchmarks.
    \item[] \textbf{Claim 3:} \label{Claim_3} Score-Max and SDM demonstrate strong aggregation capabilities in LSR, showing robustness to datasets and segment variations.
\end{itemize}
In addition, we conduct several experiments to validate the robustness of lexical representations on document retrieval. Our experiments are designed for the following research questions.
\begin{itemize}
    \item[] \textbf{RQ 1:} \label{RQ_1} Which segments in the relevant document contribute the most relevance to query-document relevance?
    \item[] \textbf{RQ 2:} \label{RQ_2}
    How do terms in first segment affect the query-document relevance? 
    \item[] \textbf{RQ 3:} \label{RQ_3} How do segment count and document length affect document-query scoring methods?
\end{itemize}
\par


In sum, this study reproduces and evaluates segment-based aggregation methods with LSR adaptation~\cite{nguyen2023adapting}, including Score-Max, ExactSDM, and SoftSDM, supporting the understanding of proximity-based relevance estimation and its scalability for processing documents. 
We also analyze the effects of segment and term frequency, targeting to enhance the robustness of retrieval models across diverse datasets and document structures, such as the scenarios of varying document lengths. Collectively, these contributions not only reinforce key findings from prior research but also broaden the potential for efficient and accurate long-document retrieval with LSR.

\section{Preliminary}

\subsection{LSR for Document Retrieval}
Learned Sparse Retrieval (LSR) methods \cite{formal2021splade,ma2019deepct} score a query-document pair by first generating sparse vectors for both the query and the document and then by calculating the dot product between them.
There are two encoders, query $f_Q$ and document encoder $f_D$. The ranking score is then obtained via dot product between query $q$ and document $d$ representations \cite{formal2021splade}, given by  $score(q,d) = f_Q(q) \cdot f_D(d) = w_q \cdot w_d = \sum_{i=1}^{|V|} w_q^i w_d^i $, where $w^i$ represent the weight of the $i^{th}$ term of the vocabulary.
These vectors are high-dimensional and sparse, with most dimensions being zero, which makes them connected to the vocabulary and closely resemble traditional sparse retrieval methods like BM25. However, unlike BM25\cite{robertson2009probabilistic}, which relies on corpus statistics such as TF-IDF, LSR learns term weights by identifying without depending on these statistics. This approach allows for the use of techniques like inverted indexing, making it efficient in terms of time and storage, as documents can be pre-encoded and integrated with existing inverted indexing libraries \cite{lin2022proposed}.

For example, Splade~\cite{formal2021splade} is a state-of-the-art bi-encoder retriever that utilizes masked language modeling (MLM) head derived from BERT \cite{devlin2018bert} to perform end-to-end term expansion and weighting. Splade encodes each term $t$ in a sequence $T$ as input into a contextualized dense representation ($h \in \mathcal{R}^{(|T|\times N)}$ with $N=768$), which is then projected into a sparse representation at the vocabulary level ($s \in \mathcal{R}^{|T|\times |V|}$ with $|V|\approx 30k$). 
The encoding process generates a logit matrix $\mathbf{W}$, , where $W_{i,j}$ is the translation score, capturing the  correlation between the $i^{th}$ term of the input sequence and the $j^{th}$ term of the vocabulary. This logit matrix can subsequently be utilized in the ranking score function introduced earlier to assess the relevance of a document in relation to a given query.
However, Splade only utilize the individual term matching. This limitation is significant for understanding phrase semantics, as documents containing phrases similar to the query should ideally be ranked higher than those with scattered similar terms \cite{tao2007exploration}.yTo address the limitation of not utilizing position embeddings, researchers have integrated the interdependence of multiple terms, like Sequential Dependence Model (SDM) \cite{metzler2005markov} , into the conventional bag-of-words retrieval models. 
.
\subsection{Sequential Dependence Model}
\label{Sequential Dependence Model}
\par
SDM assumes there is a dependence in the query terms by modeling a Markov Random Field (MRF). A joint probability over a graph $G$ is defined, whose nodes are the document random variable $D$ and the query terms $q_1,q_2,...,q_{|Q|}$.
The edges between nodes represent a dependency of them, where a node is independent of all other nodes given its neighbors. This gives the query-document conditional relevance probability:

\begin{align}
    P_\Lambda(D|Q)&=\frac{P_\Lambda(Q,D)}{P_\Lambda(Q)} \stackrel{rank}{=} P_\Lambda (Q,D) \notag
    \\ 
    P_\Lambda (Q,D)&= \frac{1}{Z_\Lambda} \notag
    \prod_{c\in C(G)} \psi (c,\Lambda)
    \stackrel{rank}{=} \sum_{c\in C(G)} \log \psi(c,\Lambda) \numberthis
    \label{first_equation}
\end{align}
In the equation, $C(G)$ is the set of cliques in the $G$ graph, $\psi (c,\Lambda)$ is a potential function parameterized by $\Lambda$, and $Z_\Lambda$ is a normalization factor.

To model sequential dependencies, the original authors decompose $\psi(c, \Lambda)$ into three potential functions:\footnote{Detailed statistical formulas are in the original paper.} (1) individual term matches ($\psi_T$), (2) exact n-gram/phrase matches ($\psi_O$), (3) exact multi-term proximity matches (ordered/ unordered within a window) ($\psi_U$).
Combining these with Eq. \eqref{first_equation}, we obtain the following ranking function:

\begin{align*}
    P(D|Q) \stackrel{rank}{=} & 
\sum_{q_i \in Q} \psi_T(q_i, D) \\ 
& + \sum_{q_i \dots q_{i+k} \in Q} \psi_O(q_i \dots q_{i+k}, D)
\\&+ \sum_{q_i \dots q_j \in Q} \psi_U(q_i \dots q_j, D) \numberthis
\end{align*}

The original work proposed an adaptation of Splade's MLM head by replacing the original functions with three new matching functions derived from query and document representations. Here, $u[q_i]$ denotes the index of $q_i$ in the vocabulary.

\begin{itemize}
    \item \textbf{Individual Term Matching:} 
        This function, analogous to LSR methods, computes the maximum similarity between a query term and all document terms, followed by a weighted sum.
        \begin{equation}
        \psi_{ST}(q_i, D) = \lambda_{ST} \max_{1 \leq r \leq |D|} w_q^i W_{r,v}^D[v_{q_i}]
        \label{Term Matching}
        \end{equation}

    \item \textbf{N-gram/Phrase Matching:} 
        This function measures the similarity between a query phrase and all document phrases, considering the importance of each term within the phrase. 
        \begin{equation}
        \psi_{SO}(q_i \dots q_{i+k}, D) = \lambda_{SO} \max_{1 \leq r \leq |D|-k} \sum_{l=0}^k w_q^{i+l} W_{r+l,v}^D[v_{q_{i+l}}]
        \label{N-gram}
        \end{equation}

    \item \textbf{Proximity Matching:} 
        This function approximates the maximum likelihood of translating a set of query terms to terms within a document window, regardless of order, while considering term importance.
        \begin{equation}
        \psi_{SU}(q_i \dots q_{i+k}, D) = \lambda_{SO} \max_{1 \leq r \leq |D|-p} \sum_{h=i}^{i+k} w_q^h 
        \left[
        \max_{r \leq l < r+p} W_{l,v}^D[v_{q_h}]
        \right]
        \label{Proximity Matching}
        \end{equation}
\end{itemize}

Based on the new equations, two models have been proposed: Soft SDM and Exact SDM. Soft SDM employs the original functions with document expansion, whereas Exact SDM does not incorporate it. Instead, Exact SDM evaluates the impact of phrase and proximity matching by limiting document terms to self-translation in the output logits, thereby modifying Equations \eqref{N-gram} and \eqref{Proximity Matching}.

\section{Reproducibility Experiments}

\subsection{Reproduced Methods}

\subsubsection{SDM Methods.} The original authors \cite{nguyen2023adapting} propose \textbf{Soft-SDM} and \textbf{Exact-SDM} as previously explained in Section \ref{Sequential Dependence Model}.

\subsubsection{Other Baselines.}
Additional methods that were used in the original paper as baselines are: 
\begin{itemize}
    \item \textbf{BM25}~\cite{robertson1994some} with SDM and RM3 approaches, based on statistical bag-of-words features.
    \item \textbf{Representation aggregation (Rep-*)}~\cite{li2023parade} approach based on learned sparse representations. Such aggregation was done by computing the relevance scores of the aggregated representation of documents.
    \item \textbf{Score aggregation (Score-*)}~\cite{li2023parade} indicates the score was aggregated among the query-segment scores.
\end{itemize}

\subsection{Experimental Setups}

\subsubsection{Datasets.}
The experiments are conducted in two long document retrieval benchmarks, the MSMARCO Document Ranking (MSDoc)~\cite{nguyen2016ms} and TREC Robust04 \cite{voorhees2003overview}. 
MSDoc contains 3.2 million documents, with 367K training queries and 5.2K dev queries. TREC Robust04 \cite{voorhees2003overview} is composed of approximately 0.5 million news articles with 250 testing queries. Each query contains three categories, including title description and narrative, which during alignment with the original work, only the description field was used. 
In addition, we access these data via \texttt{ir\_datasets}\footnote{\url{https://github.com/allenai/ir_datasets}}~\cite{macavaney2021simplified}.  

\subsubsection{Training.}
Following the original paper, we use DistilBERT\cite{sanh2020distilbert} as our encoder backbone of LSR.
The initialized checkpoint~\footnote{\url{https://huggingface.co/lsr42/qmlp_dmlm_msmarco_distil_kl_l1_0.0001}} of the bi-encoders were fine-tuned on MSMARCO Passage Ranking dataset using hard negatives and distillation from cross-encoders~\cite{reimers2019sentence}.

Particularly, we only fine-tune the aggregation parameters with the MSMARCO Document Ranking dataset and freeze both the document and query encoders.
The hyperparameters (e.g., learning rate, optimizers, etc.) are set as reported in the original repository. 
In addition, we found n-grams ($n$) and proximity $(prx)$ are two of the most sensitive parameters; thus, we also include the hyperparameter tuning in Section~\ref{Hyperparameter-tuning.}. N-grams refer to the size of the window used in Equation \eqref{N-gram}, while proximity indicates the number of tokens to be considered in Equation \eqref{Proximity Matching}.

\subsubsection{Retrieval Evaluation.}\label{Candidate List}
To evaluate ExactSDM and SoftSDM on document retrieval, we first obtained the top-$200$ document candidates by collecting them from the top-$10K$ retrieved segments. We considered both the candidate lists provided by the original authors and our reproduced ones. Our candidate list was generated using the baseline score-max method for the initial retrieval task, selecting the top-$200$ documents as candidates, following the original authors' procedures.
However, we observed in Table~\ref{tab:R@200} improved recall using our candidate list\footnote{The authors' candidate list is in their repository; ours is in our repository.} for the MSMARCO Document Ranking task; thus, we adopted our candidate list for this task. For the TREC Robust04 dataset, we observed similar recall levels and retained the original authors' candidate list.\par
We validate the retrieval effectiveness with official evaluation metrics, MRR@10 and nDCG@10, for MSDoc Dev and Robust04, respectively.
The scores are calculated by \texttt{ir\_measures}\footnote{\url{https://github.com/terrierteam/ir_measures}}~\cite{macavaney2022streamlining}. 

\begin{table}[t]

    \caption{Recall@200 Comparison on MSDoc Dev and Robust04: Reproduced vs. Provided Candidate Lists.}
    \label{tab:R@200}
    \centering
    \resizebox{\textwidth}{!}{
    \begin{tabular}{l ccccc ccccc}
        \toprule
         & \multicolumn{5}{c}{Reproduced (ours)} & \multicolumn{5}{c}{Provided} \\
         \cmidrule(lr){2-6} \cmidrule(lr){7-11}
        \#segs 
        & \expand{1} & \expand{2} & \expand{3} & \expand{4} & \expand{5} & 
        1 & 2 & 3 & 4 & 5\\
        \midrule
        MSDoc Dev
        & 91.33 & 92.72 & 93.14 & 93.30 & 93.20 
        & 90.10 & 91.28 & 91.91 & 92.20 & 92.22 \\
        \midrule
        Robust04
        & 42.67 & 46.82 & 48.71 & 49.34 & 49.98 
        & 42.67 & 46.82 & 48.71 & 49.34 & 49.98\\ 
        \bottomrule
    \end{tabular}
    }
\end{table}

\subsection{Implementation Detail}
The original code is publicly available in a GitHub \hyperlink{https://github.com/thongnt99/lsr-long}{repository by Nguyen et al. }\cite{nguyen2023adapting}.
We conducted the same set of experiments using the original hyperparameters specified in the repository by Nguyen et al. to evaluate the reproducibility of the reported results and claims. While the experiments were largely successful, a minor runtime error occurred related to locating the correct path for the Hugging Face pre-trained model. This issue did not impact the overall results and was easily resolved.
\par
The experiments were run on an AMD Rome CPU (128 threads) and an NVIDIA A100 GPU. Each CPU-based experiment took approximately one hour to complete. The GPU required around 60 hours to execute the Rep-* and Score-max methods. Training the SDM models took one hour per experiment, while the evaluation of each experiment on Robust04 required roughly 10 minutes. Given the larger-than-normal document sizes, these experiments were more memory-intensive than compute-intensive, necessitating careful selection of an optimal batch size to balance memory usage and processing speed.

\section{Results and Analysis}

\subsection{Reproduced Results}

The reproduced results are presented in Table \ref{Reproducibility table}. From this table, we observe that the reproduced outcomes closely match the original ones, with most results differing by only the second decimal place (i.e., the subscripts). However, we identified a technical inconsistency in the original results, particularly in the MSDoc experiment with four segments. We found that if certain columns are reinterpreted, the values align more closely with ours. Specifically, it appears the original authors may have inadvertently switched the columns: \textit{Score-Max} with \textit{Rep-(sc.)sum}, \textit{Rep-(sc.)sum} with \textit{Rep-Max}, and \textit{Rep-Max} with \textit{Score-Max}. When these adjustments are made, discrepancies fall within the second decimal place as well. In addition, we observed minor discrepancies of around 2\% in the \textit{SoftSDM} results on the Robust04 experiments. These discrepancies don't change the overall findings of the main authors; thus, these results have credibility to the authors' claims and help confirm the overall trend observed in their findings.

\begin{table*}[t]
    \centering
    \caption{Evaluation of Aggregation Baselines and Fine-tuned SDM Variants (ExactSDM/SoftSDM): MRR@10 on MSDoc and NDCG@10 on Robust04 (n=2, prx=8). Main values show experiment results; subscript values show difference from baseline.}
    \label{Reproducibility table}
    \resizebox{\textwidth}{!}{
    \begin{tabular}{cc cc c ccc cc}
        \toprule
        & \multicolumn{1}{c}{}& 
        \multicolumn{2}{c}{LSR w/ SDM} & 
        \multicolumn{4}{c}{LSR baselines} & 
        \multicolumn{2}{c}{BM25}\\
        \multicolumn{2}{c}{\#segs} 
        & \expand{\thead{Exact\\SDM}} 
        & \expand{\thead{Soft\\SDM}} 
        & \multicolumn{1}{c}{\expand{\thead{Score\\max}}} & \multicolumn{1}{c}{\thead{Rep\\max}} 
        & \multicolumn{1}{c}{\thead{Rep(sc.)\\sum}} 
        & \multicolumn{1}{c}{\thead{Rep(sc.)\\mean}} 
        & \multicolumn{1}{c}{\expand{\thead{BM25\\SDM}}} & \multicolumn{1}{c}{\thead{BM25\\RM3}} \\
        \midrule
        \multirow{5}{*}{\rotatebox[origin=c]{90}{\thead{MSDoc Dev}}} 
        & 1 
        & \textbf{36.99} {\tiny -0.09} 
        & 36.92 {\tiny -0.06} 
        & 36.62 { \tiny -0.01} 
        & 36.62 {\tiny -0.01} 
        & 36.62 {\tiny -0.01} 
        & 36.62 {\tiny -0.01} 
        & 25.95 {\tiny -0.14} 
        & 21.13 {\tiny 0} \\ 
        & 2 
        & \textbf{37.37} {\tiny +0.08}  
        & 37.25 {\tiny -0.28}  
        & 36.79 {\tiny -0.01}  
        & 35.15 {\tiny +0.01}  
        & 27.11 {\tiny +0.02}  
        & 31.62 {\tiny +0.04}  
        & 26.19 {\tiny +0.02}  
        & 19.69 {\tiny 0} \\
        & 3 
        & \textbf{37.34} {\tiny -0.02} 
        & 37.15 {\tiny -0.26} 
        & 36.70 {\tiny -0.02} 
        & 33.89 {\tiny -0.02} 
        & 19.70 {\tiny -0.01} 
        & 29.04 {\tiny +0.02} 
        & 25.93 {\tiny +0.02} 
        & 18.83 {\tiny 0} \\
        & 4 
        & \textbf{37.00} {\tiny -0.03} 
        & 36.85 {\tiny +0.05} 
        & 36.34 {\tiny \textbf{+3.62}} 
        & 32.70 {\tiny \textbf{+18.18}} 
        & 14.53 {\tiny \textbf{-21.83}} 
        & 27.94 {\tiny +0.00} 
        & 25.58 {\tiny -0.06}
        & 17.94 {\tiny 0} \\
        & 5
        & \textbf{36.94} {\tiny -0.01}
        & 36.80 {\tiny +0.01}
        & 36.21 {\tiny -0.03}
        & 31.96 {\tiny -0.01}
        & 12.16 {\tiny -0.01}
        & 27.52 {\tiny +0.00}
        & 25.47 {\tiny -0.01}
        & 17.34 {\tiny 0}
        \\
        \midrule
        \multirow{5}{*}{\rotatebox[origin=c]{90}{\thead{Robust04}}} 
        & 1 
        & \textbf{46.62} {\tiny -0.01} 
        & 45.23 {\tiny \textbf{-1.42}}
        & 44.82 {\tiny -0.06}
        & 44.82 {\tiny -0.06}
        & 44.82 {\tiny -0.06}
        & 44.82 {\tiny -0.06}
        & 39.70 {\tiny +0.07}
        & 42.05 {\tiny 0}\\ 
        & 2 
        & \textbf{47.92} {\tiny -0.06}
        & 45.64 {\tiny \textbf{-2.30}}
        & 45.72 {\tiny +0.01}
        & 43.98 {\tiny +0.10}
        & 42.25 {\tiny -0.02}
        & 44.33 {\tiny +0.01}
        & 42.14 {\tiny +0.10}
        & 43.04 {\tiny 0}\\
        & 3 
        & \textbf{48.60} {\tiny +0.01}
        & 45.85 {\tiny \textbf{-2.43}}
        & 46.47 {\tiny +0.05}
        & 42.58 {\tiny +0.07}
        & 36.47 {\tiny +0.04}
        & 42.71 {\tiny -0.01}
        & 42.42 {\tiny +0.40}
        & 43.59 {\tiny 0}\\
        & 4 & \textbf{48.76} {\tiny -0.11}
        & 46.23 {\tiny \textbf{-2.33}}
        & 46.57 {\tiny +0.08}
        & 41.02 {\tiny -0.05}
        & 32.17 {\tiny +0.03}
        & 42.25 {\tiny +0.00}
        & 42.83 {\tiny +0.34}
        & 43.42 {\tiny 0}
        \\
        & 5 
        & \textbf{48.60} {\tiny -0.44}
        & 46.20 {\tiny \textbf{-2.45}}
        & 46.45 {\tiny +0.08}
        & 40.58 {\tiny +0.08}
        & 28.41 {\tiny +0.00}
        & 41.74 {\tiny -0.03}
        & 43.14 {\tiny +0.68}
        & 44.02 {\tiny 0}\\
        \bottomrule
    \end{tabular}
    }
\end{table*}

\begin{table*}[t]
    \centering
    \caption{Evaluation of Aggregation Baselines and Fine-tuned SDM Variants (ExactSDM/SoftSDM): MRR@10 on MSDoc and NDCG@10 on Robust04 (n=2, prx=8). Main values show show difference from original paper; subscript values experiment results.}
    \label{Reproducibility table}
    \resizebox{\textwidth}{!}{
    \begin{tabular}{cc cc c ccc cc}
        \toprule
        & \multicolumn{1}{c}{}& 
        \multicolumn{2}{c}{LSR w/ SDM} & 
        \multicolumn{4}{c}{LSR baselines} & 
        \multicolumn{2}{c}{BM25}\\
        \multicolumn{2}{c}{\#segs} 
        & \expand{\thead{Exact\\SDM}} 
        & \expand{\thead{Soft\\SDM}} 
        & \multicolumn{1}{c}{\expand{\thead{Score\\max}}} & \multicolumn{1}{c}{\thead{Rep\\max}} 
        & \multicolumn{1}{c}{\thead{Rep(sc.)\\sum}} 
        & \multicolumn{1}{c}{\thead{Rep(sc.)\\mean}} 
        & \multicolumn{1}{c}{\expand{\thead{BM25\\SDM}}} & \multicolumn{1}{c}{\thead{BM25\\RM3}} \\
        \midrule
        \multirow{5}{*}{\rotatebox[origin=c]{90}{\thead{MSDoc Dev}}} 
        & 1 
        & {\tiny \textbf{36.99} } -0.09
        & {\tiny 36.92 } -0.06
        & {\tiny 36.62 } -0.01
        & {\tiny 36.62 }-0.01 
        & {\tiny 36.62 } -0.01
        & {\tiny 36.62 } -0.01
        & {\tiny 25.95 } -0.14
        & {\tiny 21.13 } 0\\ 
        & 2 
        & {\tiny \textbf{37.37} } +0.08
        & {\tiny 37.25 } -0.28
        & {\tiny 36.79 } -0.01
        & {\tiny 35.15 } +0.01
        & {\tiny 27.11 } +0.02
        & {\tiny 31.62 } +0.04
        & {\tiny 26.19 } +0.02
        & {\tiny 19.69 } 0\\
        & 3 
        & {\tiny \textbf{37.34} } -0.02
        & {\tiny 37.15 } -0.26
        & {\tiny 36.70 } -0.02
        & {\tiny 33.89 } -0.02
        & {\tiny 19.70 } -0.01
        & {\tiny 29.04 } +0.02
        & {\tiny 25.93 } +0.02
        & {\tiny 18.83 } 0\\
        & 4 
        & {\tiny \textbf{37.00} } -0.03
        & {\tiny 36.85 } +0.05
        & {\tiny 36.34 } \textbf{+3.62}
        & {\tiny 32.70 } \textbf{+18.18}
        & {\tiny 14.53 } \textbf{-21.83}
        & {\tiny 27.94 } +0.00
        & {\tiny 25.58 } -0.06
        & {\tiny 17.94 } 0\\
        & 5
        & {\tiny \textbf{36.94} } -0.01
        & {\tiny 36.80 } +0.01
        & {\tiny 36.21 } -0.03
        & {\tiny 31.96 } -0.01
        & {\tiny 12.16 } -0.01
        & {\tiny 27.52 } +0.00
        & {\tiny 25.47 } -0.01
        & {\tiny 17.34 } 0
        \\
        \midrule
        \multirow{5}{*}{\rotatebox[origin=c]{90}{\thead{Robust04}}} 
        & 1 
        & {\tiny \textbf{46.62} } -0.01
        & {\tiny45.23 } \textbf{-1.42}
        & {\tiny44.82 } -0.06
        & {\tiny44.82 } -0.06
        & {\tiny44.82 } -0.06
        & {\tiny44.82 } -0.06
        & {\tiny39.70 } +0.07
        & {\tiny42.05 } 0\\ 
        & 2 
        & {\tiny\textbf{47.92} } -0.06
        & {\tiny45.64 } \textbf{-2.30}
        & {\tiny45.72 } +0.01
        & {\tiny43.98 } +0.10
        & {\tiny42.25 } -0.02
        & {\tiny44.33 } +0.01
        & {\tiny42.14 } +0.10
        & {\tiny43.04 } 0\\
        & 3 
        & {\tiny\textbf{48.60} } +0.01
        & {\tiny45.85 }\textbf{-2.43}
        & {\tiny46.47 } +0.05
        & {\tiny42.58 } +0.07
        & {\tiny36.47 } +0.04
        & {\tiny42.71 } -0.01
        & {\tiny42.42 } +0.40
        & {\tiny43.59 } 0\\
        & 4 & {\tiny\textbf{48.76} } -0.11
        & {\tiny46.23 } \textbf{-2.33}
        & {\tiny46.57 } +0.08
        & {\tiny41.02 } -0.05
        & {\tiny32.17 } +0.03
        & {\tiny42.25 } +0.00
        & {\tiny42.83 } +0.34›
        & {\tiny43.42 } 0
        \\
        & 5 
        & {\tiny \textbf{48.60}} -0.44
        & {\tiny 46.20} \textbf{-2.45}
        & {\tiny 46.45} +0.08
        & {\tiny 40.58} +0.08
        & {\tiny 28.41}+0.00
        & {\tiny 41.74} -0.03
        & {\tiny 43.14} +0.68
        & {\tiny 44.02} 0\\
        \bottomrule
    \end{tabular}
    }
\end{table*}

\subsubsection{\hyperref[Claim_1]{\textbf{Claim 1:}} \textbf{As the number of segments increases, the performance of representation aggregation decreases, whereas Score-Max remains unaffected.}}
In evaluating the effectiveness of Rep-* and Score-*, a distinct performance emerges, highlighting the differences in relevance ranking and the impact of segment count. Specifically, \textit{Rep(sc.)-sum} exhibits a lower MRR on the MSDoc than \textit{BM25-RM3}, suggesting a tendency to favor longer documents over those with higher relevance. The \textit{Rep(sc.)-mean} method partially mitigates this issue by normalizing for segment count. In comparisons between \textit{Rep(sc.)-mean} and \textit{Rep-max}, \textit{Rep(sc.)-max} outperforms the MSDoc dataset, whereas \textit{Rep(sc.)-mean} achieves the better performance on the Robust04. Nevertheless, the performance decrease is observed across all methods as segment count increases.
\par
Among the baseline approaches, \textit{Score-max} consistently demonstrates the best performance. On the MSDoc dataset, its MRR@10 experiences only a slight decline, remaining above 36\%, while other methods drop below this level. On the Robust04 dataset, \textit{Score-max} achieves an increase up to 46.57\%, contrasting with declines observed in other methods. Unlike \textit{Rep-max}, which prioritizes global term importance, \textit{Score-max} evaluates each segment independently, thereby reducing noise accumulation and enhancing robustness. These findings support \hyperref[Claim_1]{Claim 1}.

\subsubsection{\hyperref[Claim_2]{\textbf{Claim 2:}}\textbf{The ExactSDM and SoftSDM models improve LSR performance
through proximity-based matching, with ExactSDM emerging as the most
effective aggregation method for LSR.}} 
The declining performance of Score-max highlights the importance of proximity scoring for the effective evaluation of long documents using LSR. As shown in Table \ref{Reproducibility table}, ExactSDM and SoftSDM, with appropriate hyperparameter tuning, outperform all baseline methods, especially as segments increase. For instance, while using ExactSDM, the performance improves until two additional segments are added, followed by a slight decrease with further segmentation. An impact assessment has been conducted to explore how segment addition affects aggregation methods, and this will be discussed in a later section.
In comparison to Score-max, both SDM variants achieve higher MRR@10, with improvements of approximately 1\% on MSDoc Document. On Robust04 (zero-shot), improvements range from 1.8\% to 2.13\%. Notably, SDM variants sustain slight performance gains even as segments increase, denoting their capability to effectively utilize longer contexts. These findings support \hyperref[Claim_2]{Claim 2}.

 \begin{table*}[t]
    \caption{
    Evaluation of Fine-tuned SDM Variants (ExactSDM and SoftSDM): MRR@10 on MSDoc and NDCG@10 on Robust04 across different hyperparameters.
}
    \label{Hyperparameters results}
    \centering
    \begin{tabular}{cc cc cc }
        \toprule
        \multicolumn{2}{c}{} & \multicolumn{2}{c}{\thead{ExactSDM}} 
        & \multicolumn{2}{c}{\thead{SoftSDM}}\\ 
        \multicolumn{2}{c}{\thead{\#segs.} }
        & \multicolumn{1}{c}{\expand{\thead{$n=2$\\$prx=8$}}} 
        & \multicolumn{1}{c}{\expand{\thead{$n=5$\\$prx=10$}}} 
        & \multicolumn{1}{c}{\expand{\thead{$n=2$\\$prx=8$}}} 
        & \multicolumn{1}{c}{\expand{\thead{$n=5$\\$prx=10$}}} \\
        \midrule
        \multirow{5}{*}{\rotatebox[origin=c]{90}{\thead{MSDoc Dev}}} 
        & 1 & 36.99 & 36.80 & 36.92 & \textbf{37.01}\\ 
        & 2 & \textbf{37.37} & 36.51  & 37.25 & 37.27 \\
        & 3 & \textbf{37.34} & 36.55  &  37.15 & 37.07 \\
        & 4 & \textbf{37.00} & 36.42  & 36.85 & 36.76 \\
        & 5 & \textbf{36.94} & 36.28 & 36.80 & 36.65 \\
        \midrule
        \multirow{5}{*}{\rotatebox[origin=c]{90}{\thead{Robust04}}} 
        & 1 & \textbf{46.62}  & 45.06  & 45.23 & 46.87 \\ 
        & 2 & \textbf{47.92} & 45.68  & 45.64 & 47.86 \\
        & 3 & \textbf{48.60} & 46.54 & 45.85 & 48.08 \\
        & 4 & \textbf{48.76} & 46.60 & 46.23 & 48.42 \\
        & 5 & \textbf{48.60} & 46.45 & 46.20 & 48.65 \\
        \bottomrule
    \end{tabular}
\end{table*}

\subsubsection{\hyperref[Claim_3]{\textbf{Claim 3:}}\textbf{
Score-Max and SDM demonstrate strong aggregation capabilities in LSR, showing robustness to datasets and segment variations.}} 
The results indicate that both \textit{Score-max} and \textit{SDM} effectively aggregate information, maintaining performance stability in different datasets. Among the baseline methods, \textit{Score-max} consistently shows the most robust overall performance. On the MSDoc, its MRR@10 decreases only slightly, remaining within a single percentage point. On the Robust04 dataset, \textit{Score-max} increases to 46.57 by the fourth segment before stabilizing with a minimal decline. Additionally, both \textit{ExactSDM} and \textit{SoftSDM} exhibit only minor declines in MRR@10, with no drop exceeding a single percentage point on MSDoc. Both methods also demonstrate consistent improvement on Robust04, underscoring their robustness. These results provide strong support for \hyperref[Claim_3]{Claim 3}.

\label{Hyperparameter-tuning.}
In addition to the proposed claims, the authors emphasize that hyperparameter tuning is essential for optimizing \textit{SDM} methods. The advice is to increase the $n$-gram and proximity ($prx$) to enhance the nDCG@10 score for Robust, even though this may slightly reduce the MRR@10 for MSDoc.  To investigate this further, we conducted two additional experiments, namely n-grams of 2 and proximity of 8 and n-grams of 5 and proximity of 8. 
Our experiments indicate a performance boost of 1\% to 2\% on Robust04 for \textit{SoftSDM} with these adjustments. In contrast, we could not replicate similar gains for \textit{ExactSDM}, highlighting the need for further research in this area. Future tuning efforts could benefit from insights provided by prior research in the field \cite{bendersky2010learning,metzler2005markov}.

\subsection{Relevance Variation of Document Segments}
\label{Document segment influence on relevance}
To better understand the influence of each document segment and answer the \hyperref[RQ_1]{RQ 1}, an impact analysis was conducted. We applied max-score in trec-robust dataset to identify the highest-scoring segment for each document-query pair. The analysis was performed under two scenarios: one including both relevant and irrelevant document-query pairs and the other focusing solely on relevant pairs.

\begin{figure}[t]
\ffigbox{%
\begin{subfloatrow}
\ffigbox{%
  \includegraphics[width=\linewidth]{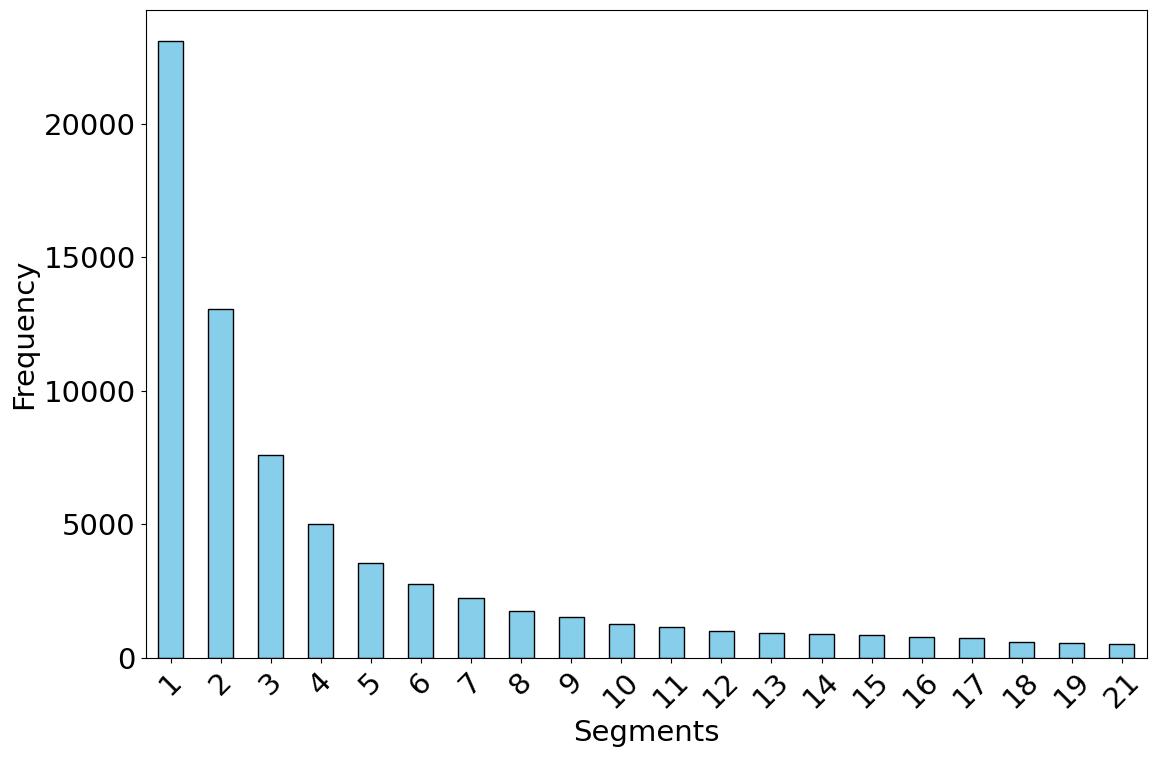}
}{%
  \caption{The highest scoring segments for relevant and irrelevant document-query pairs} \label{1a}%
}
\ffigbox{%
  \includegraphics[width=\linewidth]{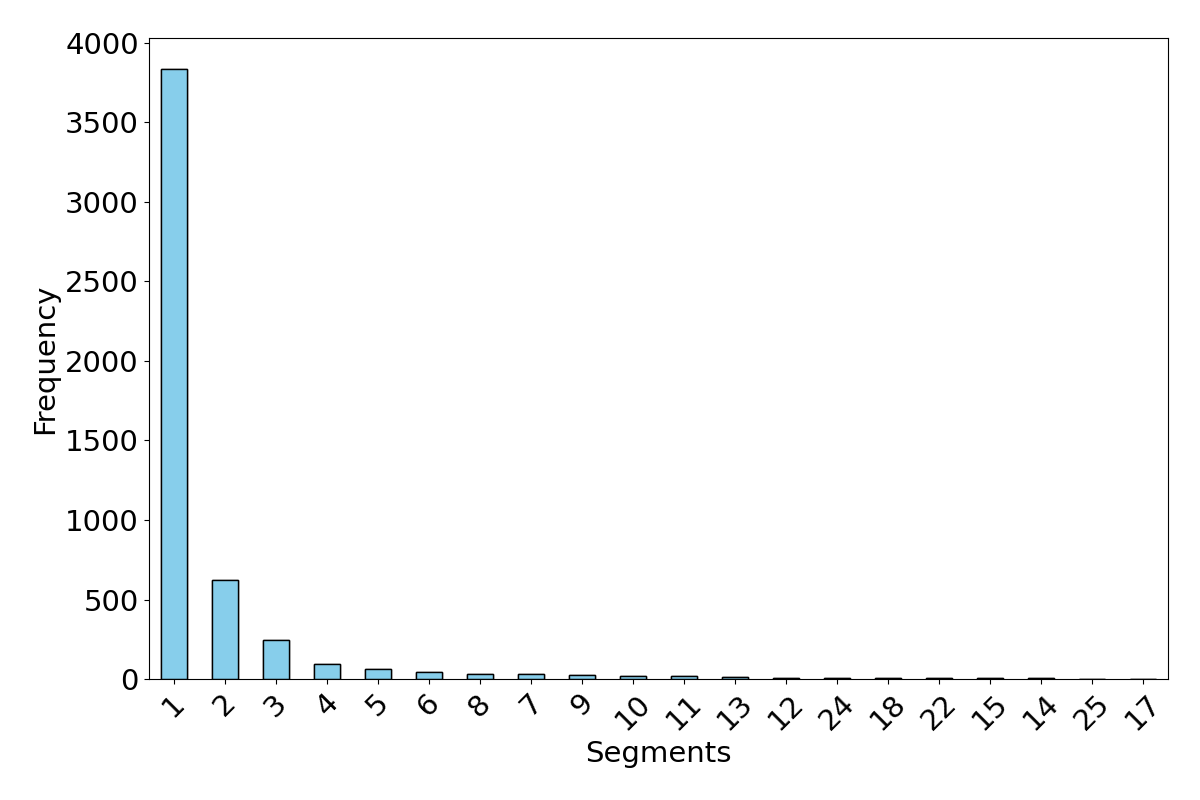}
}{%
  \caption{The highest scoring segments for only relevant document-query pairs.} \label{1b}%
}
\end{subfloatrow}
}{\caption{Analysis of Document Segment Relevance Variation: Segment distribution vs. frequency.} \label{our_results0} }
\end{figure}

As shown in Fig. \ref{1a}, for the first scenario, the first three segments consistently demonstrate the highest influence, with the remaining segments displaying a relatively uniform distribution. In the second scenario, indicated by Fig. \ref{1b}, the first three segments again show the highest importance, but the influence of the remaining segments is less uniform. In both cases, we conclude that the initial segments are the most influential. Focusing primarily on these segments can yield effective results, achieving approximately 36\% MMR@10 on MSDoc, as observed in Table \ref{Reproducibility table}.

\subsection{Dependence of Sparse Segment Representation}
Identifying which document segment has the most influence on scoring is important but does not fully reveal the dependencies between segments. Understanding these dependencies is crucial for a comprehensive view of each segment's contribution to the overall score. To address this, we conducted an effect study on the sparse representations of each document segment using the Score-max setting in the trec-robust dataset to investigate \hyperref[RQ_2]{RQ 2}. From the results of Section \ref{Document segment influence on relevance}, we found that the first segment is important to the relevance score of the document. Thus, this analysis focuses on the dependence of the first segment relative to others, specifically examining interactions with the second, third, fourth, and sixth segments. 
To examine these scoring dependencies, we categorized terms into three types: 
\begin{enumerate} 
    \item \textit{Unique terms}: Terms appearing exclusively in one segment, specifically the first segment. 
    \item \textit{Intersection terms}: Terms shared between first segment into other segments.
    \item \textit{Global terms}: Terms appearing across first and multiple other segments.
\end{enumerate}

\begin{figure}[t]
\ffigbox[0.99\textwidth]{%
    
    \begin{subfloatrow}
    \ffigbox{
        \includegraphics[width=0.95\linewidth]{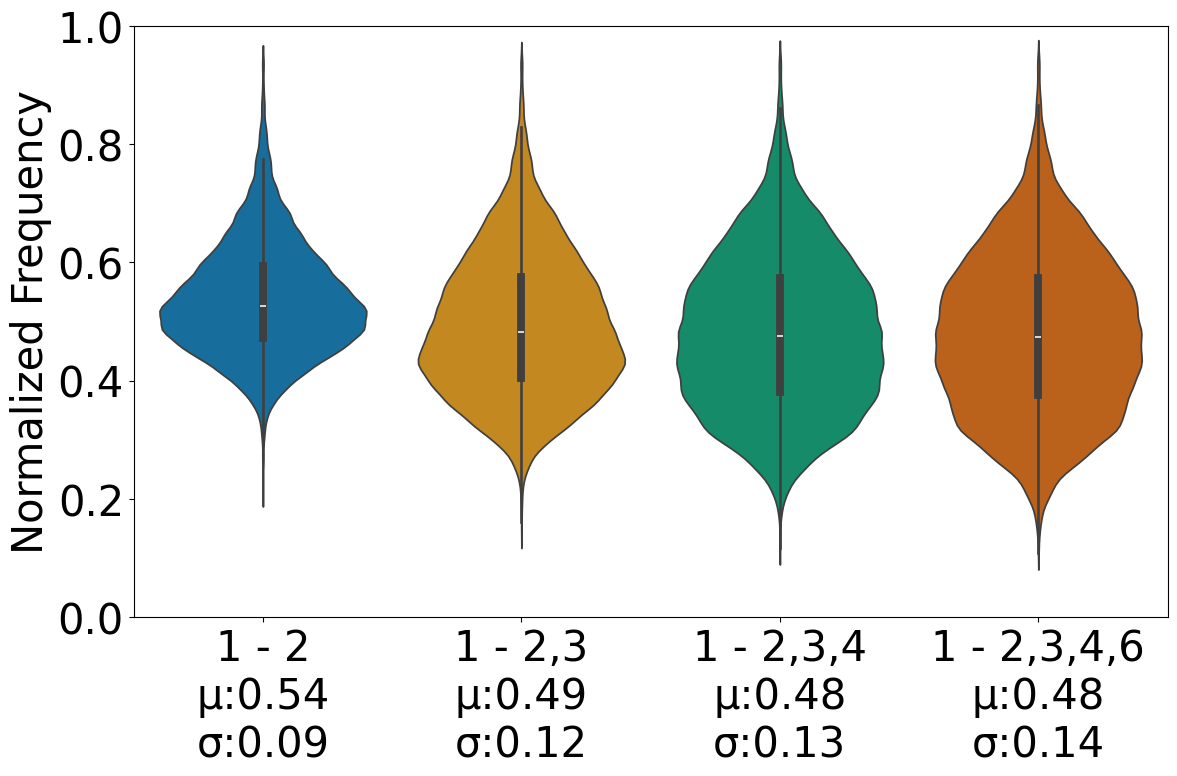}
    }{%
        \caption{Unique terms of 1st segment across other segments.} \label{2c}%
    }

    \ffigbox{
        \includegraphics[width=0.95\linewidth]{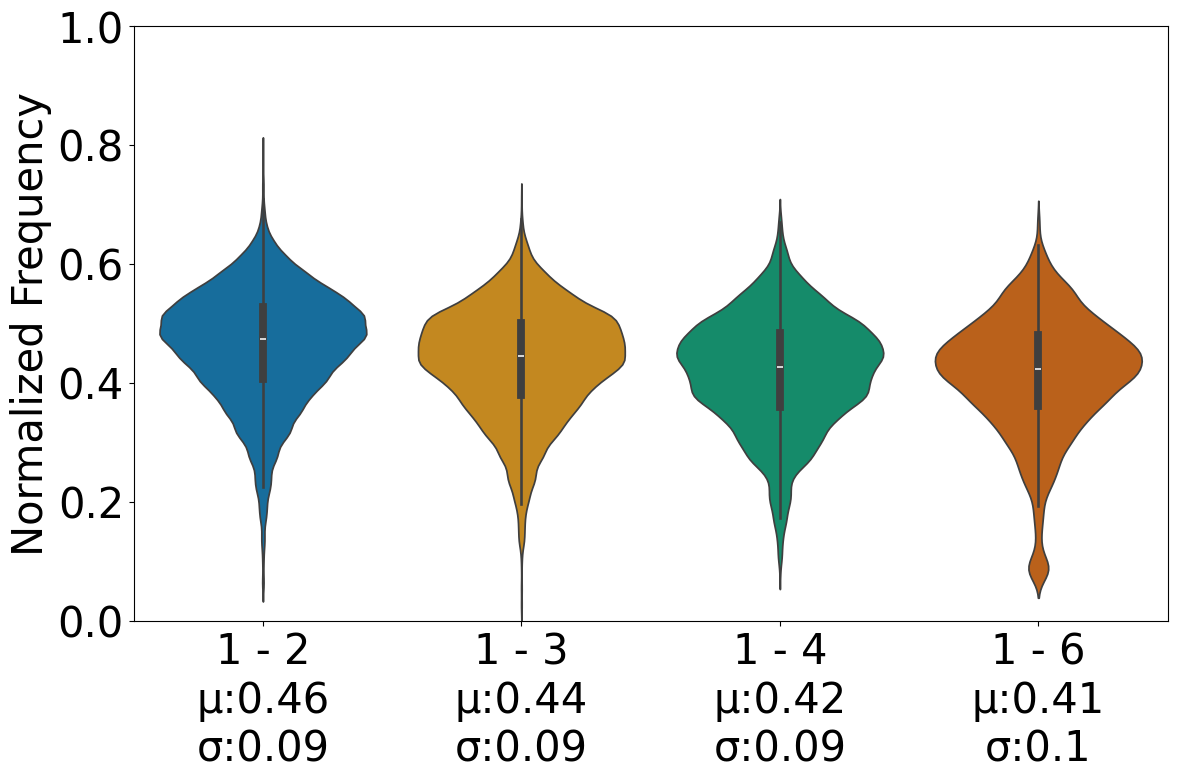}
    }{%
        \caption{Intersection terms of 1st segment with other segments.} \label{2a}%
    }
    
    \end{subfloatrow}
}

\ffigbox[0.99\textwidth]{
    \begin{subfloatrow}
    \ffigbox{
        \includegraphics[width=0.95\linewidth]{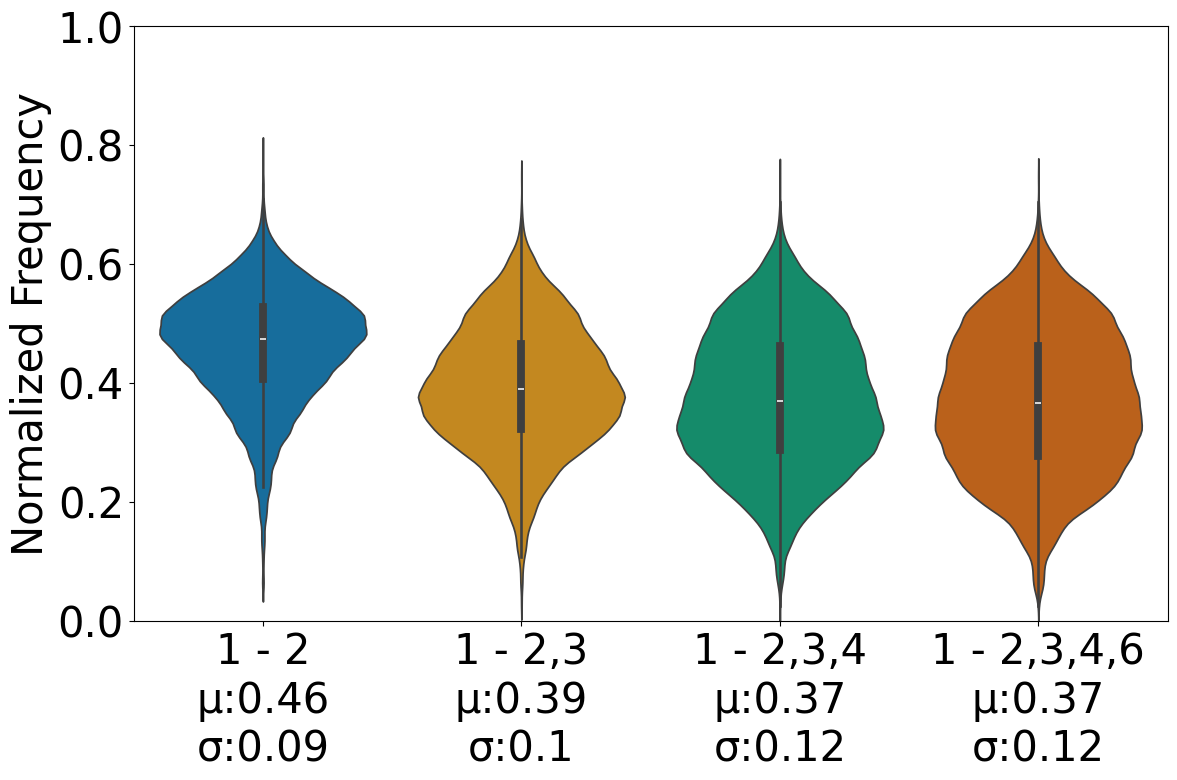}
    }{
        \caption{Global terms of 1st segment across multiple segments.} \label{2b}%
    }
    
    \ffigbox{%
        \includegraphics[width=0.95\linewidth]{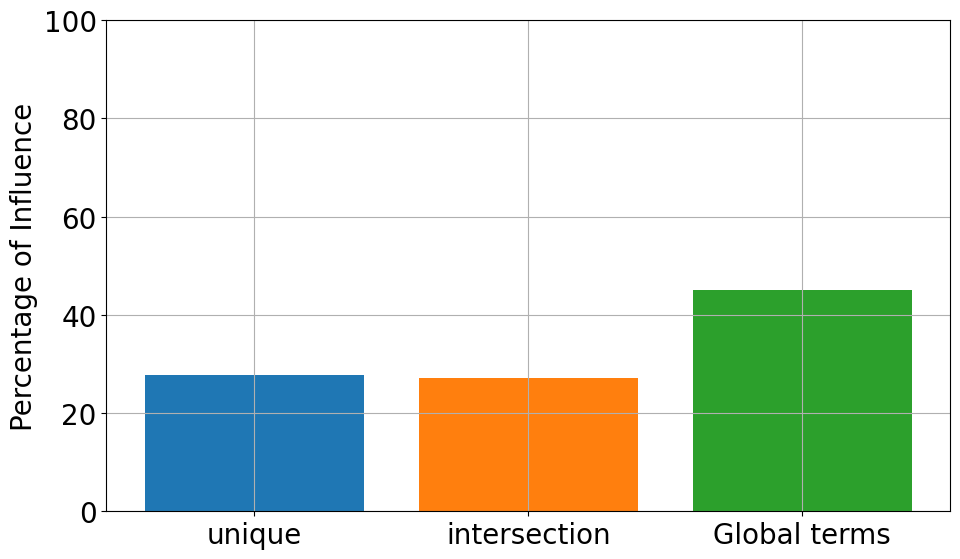}
    }{%
        \caption{The influence percentages of unique, intersection, and global terms on the query's score for the first segment.} \label{2d}%
    }
  \end{subfloatrow}
}{
\caption{Sparse Segment Representation Analysis: (a)-(c) Distributions of unique/intersection/global terms of 1st segment; (d) The influence percentages of unique, intersection, and global terms on the query's score for the first segment.} \label{fig:Dependence of Sparse Segment} }

\end{figure}

In the first category, Unique Terms, we analyzed terms appearing exclusively in the first segment. Fig. \ref{2a} shows that as we increase the number of segments, the proportion of unique terms stabilizes at a mean of $0.48$, indicating that in segment one there are significantly number of unique terms compared to other types.
In the second category, Intersection Terms, we examined terms appearing in pairs of segments. Fig. \ref{2b} reveals that segments closer to each other (e.g., 1-2) share more terms than distant segments (e.g., 1-6), with the mean dropping from $0.46$ to $0.41$. 
In the third category, Global Terms, we studied terms present across all segments. Fig. \ref{2c} shows that increasing the number of segments does not significantly decrease the total global terms. The distributions remain notably similar when comparing segments 2,3 with 2,3,4 and 2,3,4,6, suggesting that the global terms of the document stabilize after a certain number of segments.

From Figure \ref{2d}, we can evaluate the impact of unique, intersection, and global terms on the query-document score, based on the Score-Max of the first segment. The data indicates that the unique terms in the first segment significantly influence the document-query score, accounting for approximately $30\%$. Furthermore, global terms are more impactful than intersection terms, nearly doubling their percentage and achieving over $40\%$. In contrast, intersection terms are less significant than both unique and global terms, as their percentage is lower. Consequently, unique and global terms play crucial roles in determining the final score.

\subsection{Evaluating SoftSDM and ExactSDM Across Document Lengths} 

\begin{table*}[t]
    \centering
    \caption{Statistical Overview of MSMARCO Document Developer Dataset (Original and Modified): Total item counts and percentages per file.}
    \label{tab:statistics}
    \begin{tabular}{c|c|ccccc}
        \toprule
        \multicolumn{1}{c}{\textbf{Dataset}} & \multicolumn{1}{c}{\textbf{Queries}}& \multicolumn{5}{c}{\textbf{Number Candidate Document}}\\
        \cmidrule(lr){3-7}
        \multicolumn{1}{c}{}& \multicolumn{1}{c}{} &\multicolumn{1}{c}{\#segs$\leq$1}&\#segs$\leq$2&\#segs$\leq$3&\#segs$\leq$4&\#segs$\leq$5
        \\
        \midrule
        \thead{MSDoc Dev\\ original} & \thead{5,193\\100\%} & \thead{1,038,600} & \thead{1,038,600} & \thead{1,038,600} & \thead{1,038,600} & \thead{1,038,600}\\
        \thead{MSDoc Dev\\ short} & \thead{2,398\\0.45\%} & \thead{239,063} & \thead{208,951} & N/A & N/A & N/A\\
        \thead{MSDoc Dev\\ long} & \thead{2,795\\0.55\%} & \thead{313,114} & \thead{347,897} & \thead{379,171} & \thead{393,378} & \thead{401,854}\\
        \bottomrule
        \end{tabular}
\end{table*}

\begin{table*}[htb]
    \centering
    \caption{SDM Variants' MRR@10 Performance on MSMARCO Document Subsets (Short: $\leq2$ segments, Long: $\ge3$ segments, Original: complete dataset). Main values show experiment results; subscript values indicate difference from Score-max baseline.}
    \label{tab:altered_dataset}
    \begin{tabular}{cc 
    cc c}
        \toprule
        Dataset &  \multicolumn{1}{c}{\thead{\#segs}}
        & \multicolumn{1}{c}{\thead{ExactSDM}} & \multicolumn{1}{c}{\thead{SoftSDM}} & \multicolumn{1}{c}{\thead{Score-max}} \\
        \midrule
        \multirow{3}{*}{\rotatebox[origin=c]{90}{\thead{MSDoc\\Dev\\short}}} 
        &  1 
        & \textbf{52.57} {\tiny 0.48} 
        & 52.13 {\tiny 0.04}
        & 52.09 {\tiny 0}  \\
        &  2
        & \textbf{52.97} {\tiny 0.76} 
        & 52.54 {\tiny 0.33}
        & 52.21 {\tiny 0} \\
        \\
        \midrule
        \multirow{5}{*}{\rotatebox[origin=c]{90}{\thead{MSDoc\\Dev\\long}}} 
        &  1 
        & 41.17 {\tiny 0.44} 
        & \textbf{41.49} {\tiny 0.76} 
        & 40.73 {\tiny 0} \\
        &  2 
        & 42.23 {\tiny 0.80} 
        & \textbf{42.27} {\tiny 0.84} 
        & 41.43 {\tiny 0}  \\ 
        &  3 
        & \textbf{42.21} {\tiny 0.83} 
        & 42.07 {\tiny 0.72}
        & 41.35 {\tiny 0} \\ 
        &  4 
        & \textbf{41.72} {\tiny 0.90} 
        & 41.54 {\tiny 0.72}
        & 40.82 {\tiny 0}  \\ 
        &  5 
        & \textbf{41.62} {\tiny 1.14} 
        & 41.35 {\tiny 0.72} 
        & 40.48 {\tiny 0} \\ 
        \midrule
        \multirow{5}{*}{\rotatebox[origin=c]{90}{\thead{MSDoc Dev\\original}}} 
         & 1 
        & \textbf{36.99} {\tiny 0.37}
        & 36.92 {\tiny 0.30}
        & 36.62 {\tiny 0}  \\
        & 2 
        & \textbf{37.37} {\tiny 0.58} 
        & 37.27 {\tiny 0.48}
        & 36.79 {\tiny 0}  \\ 
        & 3 
        & \textbf{37.34} {\tiny 0.64} 
        & 37.15 {\tiny 0.45}
        & 36.70 {\tiny 0}  \\ 
        & 4 
        & \textbf{37.00} {\tiny 0.66}
        & 36.85 {\tiny 0.51}
        & 36.34 {\tiny 0}  \\ 
        & 5 
        & \textbf{36.94} {\tiny 0.73} 
        & 36.80 {\tiny 0.59}
        & 36.21 {\tiny 0}  \\ 
        \bottomrule
    \end{tabular}
\end{table*}

Based on the Table \ref{Reproducibility table}, single-segment methods, such as Score-Max and SDM variants, can be influenced by number of segments considered. To investigate this further, we conducted an ablation study and tried to answer the \hyperref[RQ_3]{RQ 3}, thus creating two new datasets from the MS MARCO Document dataset:  the short and long MSDoc Dev sets. The short dataset includes qrels where the relevant document contains at most two segments, while the long dataset includes documents with three or more segments.

To split the dataset, we first identified documents containing at most two segments (short documents) and those with more than two segments (long documents). We then split the qrels based on this split, identifying queries associated with either short or long documents. As the current task involves re-ranking, we filtered the candidate list to exclude all queries that identified short documents as relevant. From the remaining queries, we also removed any short documents. The statistics for the new dataset can be found in Table \ref{tab:statistics}.

The results reveal that SoftSDM and ExactSDM exhibit distinct trends depending on document length. As shown in Table \ref{tab:altered_dataset}, ExactSDM outperforms the baseline Score-Max model on short documents in the MSDoc Dev set, with MM@10 improvements of 0.48 and 0.76. Conversely, SoftSDM achieves higher gains on long documents, showing an increase of up to 0.84 percentage points when handling two-segment documents. However, as the segment count rises, ExactSDM’s performance surpasses that of SoftSDM. This trend reflects SoftSDM’s design, which applies document expansion. In shorter documents, this can introduce unnecessary noise, giving ExactSDM an advantage. However, for longer documents that lack sufficient segment information when the segment count is low, SoftSDM’s expansion is advantageous, as it can help compensate for the sparse content. Ultimately, when documents of varied lengths are considered, ExactSDM demonstrates better differentiation between short and long documents.

\section{Conclusion}
    
In this study, we reproduced the findings of the original paper by Nguyen et al. \cite{nguyen2023adapting} and conducted impact analysis to further validate the authors’ claims. Our results indicate that the Score-Max consistently outperforms BM25 and other aggregation techniques. Among the SDM adaptations, both SoftSDM and ExactSDM proved to be highly effective, with only minor performance differences between reported and reproduced. Our experiments also highlight the need for further tuning of hyperparameters.
Additionally, we found that the first segment play an influential role in determining document relevance, with unique terms in each segment proving just as important as global terms. Finally, we note that ExactSDM outperforms SoftSDM when compering to long document with significant segments. 
This reproducibility also highlights the importance of unique and global terms, suggesting the exploration of their full potential. 

\bibliographystyle{splncs04}
\bibliography{mybibliography}

\end{document}